\documentclass[12pt]{amsart}
\usepackage{amsfonts,amssymb}
\RequirePackage{ifpdf}
\usepackage{amsmath}
\usepackage{color}
\usepackage{pstcol,pst-node}
\usepackage{pst-plot}
\def\appendix#1{
\addtocounter{section}{1} \setcounter{equation}{0}
\renewcommand{\thesection}{\Alph{section}}
\section*{Appendix \thesection\protect\indent\quad
#1}
}
\renewcommand{\theequation}{\thesection.\arabic{equation}}

\catcode`\@=11
\def\marginnote#1{}

\newcount\hour
\newcount\minute
\newtoks\amorpm
\hour=\time\divide\hour by60 \minute=\time{\multiply\hour by60
\global\advance\minute by-\hour}
\edef\standardtime{{\ifnum\hour<12 \global\amorpm={am}%
        \else\global\amorpm={pm}\advance\hour by-12 \fi
        \ifnum\hour=0 \hour=12 \fi
        \number\hour:\ifnum\minute<10 0\fi\number\minute\the\amorpm}}
\edef\militarytime{\number\hour:\ifnum\minute<100\fi\number\minute}

\newcommand{\tcr}{\textcolor{red}}
\newcommand{\tcb}{\textcolor{blue}}
\def\draftlabel#1{{\@bsphack\if@filesw {\let\thepage\relax
      \xdef\@gtempa{\write\@auxout{\string
          \newlabel{#1}{{\@currentlabel}{\thepage}}}}}\@gtempa \if@nobreak
    \ifvmode\nobreak\fi\fi\fi\@esphack} \gdef\@eqnlabel{#1}}
    \def\@eqnlabel{}
\def\@vacuum{}
\def\draftmarginnote#1{\marginpar{\raggedright\scriptsize\tt#1}}

\def\draft{
  \oddsidemargin -.5truein
  \def\@oddfoot{\footnotesize \sl preliminary draft \hfil
    \rm\thepage\hfil\sl\today\quad\militarytime}
  \let\@evenfoot\@oddfoot \overfullrule 3pt
    \let\label=\draftlabel
    \let\marginnote=\draftmarginnote
  \def\@eqnnum{(\theequation)\rlap{\kern\marginparsep\tt\@eqnlabel}%
    \global\let\@eqnlabel\@vacuum}

  }
\voffset= - 0.5in \hoffset= - 1.0in

\newcommand{\tr}{\,{\rm Tr}\,}

\def\be{\begin{equation}}
\def\ee{\end{equation}}
\def\bea{\begin{eqnarray}}
\def\eea{\end{eqnarray}}
\def\<{\langle}
\def\>{\rangle}

\def\tr{\mathop{\rm{tr}}}

\def\ocomma{{\phantom{\Bigm|}^{\phantom {X}}_{\raise-1.5pt\hbox{,}}\!\!\!\!\!\!\otimes}}

\newtheorem{theorem}{Theorem}[section]
\newtheorem{lemma}[theorem]{Lemma}

\theoremstyle{definition}
\newtheorem{definition}[theorem]{Definition}

\newtheorem{remark}[theorem]{Remark}

\allowdisplaybreaks

\long\def\rem#1{}

\begin{document}

\title[Symplectic structures on Teichm\"uller spaces $\mathfrak T_{g,s,n}$ and cluster algebras]
{Symplectic structures on Teichm\"uller spaces $\mathfrak T_{g,s,n}$ and cluster algebras}
%{Fenchel--Nielsen symplectic forms, Goldman brackets and cluster algebras}
\author{Leonid O. Chekhov$^{\ast}$}\thanks{$^{\ast}$Steklov Mathematical
Institute, Moscow, Russia, and Michigan State University, East Lansing, USA. Email: chekhov@mi-ras.ru.\\ The work was supported in part by the RFBR grant No. 18-01-00460.}

\begin{abstract}
We recall the fat-graph description of Riemann surfaces $\Sigma_{g,s,n}$ and the corresponding Teichm\"uller spaces $\mathfrak T_{g,s,n}$  with $s>0$ holes and $n>0$ bordered cusps in the hyperbolic geometry setting. If $n>0$, we have a bijection between the set of Thurston shear coordinates and Penner's $\lambda$-lengths and we can induce, on the one hand, the Poisson bracket on $\lambda$-lengths from the Poisson bracket on shear coordinates introduced by V.V.Fock in 1997 and, on the other hand, a symplectic structure $\Omega_{\text{WP}}$ on the set of extended shear coordinates from Penner's symplectic structure on $\lambda$-lengths. We derive $\Omega_{\text{WP}}$, which turns out to be similar to the Kontsevich symplectic structure for $\psi$-classes in complex-analytic geometry, and demonstrate that it is indeed inverse to the Fock Poisson structure.
\end{abstract}

\maketitle

{{}\hfill
{\it To honour my teacher, Andrei Alekseevich Slavnov.}}

\section{Introduction}\label{s:intro}
\setcounter{equation}{0}

Studies of Poisson and symplectic structures on moduli spaces of Riemann surfaces intertwine mathematics and physics. On a theoretical physics side, these structures describe low-dimensional gravity models, whereas in mathematical physics they describe monodromies of Fuchsian systems and solutions of Painlev\'e equations. Theoretical physics methods proved to be pivotal for exploring these structures; notably, Goldman in 1986 \cite{Gold} derived his celebrated bracket on the set of geodesic functions using canonical Poisson bracket originated from the Chern--Simons theory. All this indicates the importance of finding Darboux coordinates in the corresponding Teichm\"uller spaces (spaces parameterizing equivalence classes of locally conformally equivalent Riemann surfaces).

Darboux coordinates  for moduli spaces of Riemann surfaces with holes (without bordered cusps) were identified in \cite{ChF1} with the shear coordinates for an ideal triangle decomposition obtained in  \cite{Fock1} by generalising the results of \cite{Penn1} for punctured Riemann surfaces where another set of coordinates, namely, Penner's $\lambda$-lengths, was introduced. 

A main algebraic object of studies, {\em geodesic functions} $G_\gamma = 2\cosh (\ell_\gamma/2)$ with $\ell_\gamma$ being lengths of closed geodesics, were expressed in shear coordinates of decorated Teichm\"uller spaces for Riemann surfaces with holes in \cite{ChF2}: it was shown there that all geodesic  functions are Laurent polynomials of exponentiated shear coordinates $e^{Z_\alpha/2}$ with positive integer coefficients. In  \cite{ChSh}, a general combinatorial description of geodesic  functions in terms of shear coordinates for Riemann surfaces with holes and (possibly) orbifold points was presented giving rise to new generalised cluster transformations (cluster algebras with coefficients).

The main statement of  \cite{ChF2} is that the constant Poisson bracket on the set of shear coordinates induces the Goldman bracket on the set of geodesic functions.

We do not address quantization issues in this paper restricting ourself to Poisson structures. Note however that quantum Riemann surfaces are algebras of quantum observables represented by quantum geodesic  functions. The  shear coordinates were quantized in \cite{ChF1} and in the Liouville-type parameterisation in \cite{Kashaev}, whereas quantum  flip morphisms that satisfy the quantum pentagon identity were based on the quantum dilogarithm function \cite{Faddeev}.

Shear coordinates can be identified with the $Y$-type cluster variables \cite{FZ},~\cite{FZ2}. In \cite{ChMaz2}, the quantum bordered Riemann surfaces $\Sigma_{g,s,n}$ and the corresponding quantum Teichm\"uller spaces $\mathfrak T^\hbar_{g,s,n}$ were constructed; because in the case of $\Sigma_{g,s,n}$ with $n>0$ we have a bijection and monoidal transformations between the extended shear coordinates and $\lambda$-lengths of {\em arcs} (sides of ideal triangles of an ideal triangle decomposition dual to the fat graph determining the shear coordinates), Poisson and quantum relations for $\lambda$-lengths follow easily; they appear to be \cite{ChMaz2} respectively Poisson and quantum cluster algebras introduced by Berenstein and Zelevinsky \cite{BerZel}, and since $\lambda$-lengths of arcs enjoy Ptolemy relations  \cite{Penn1}, the correspondence to cluster algebras is immediate.

The above bijection also allows deriving the symplectic structure on the set of extended shear coordinates out of the Penner symplectic structure on the set of $\lambda$-lengths. This is the main result of this paper (Theorem~\ref{th:Omega}). The obtained structure is equivalent to Papadopoulos and Penner symplectic structure \cite{PP93} on the set of $h$-coordinates; we also demonstrate that this structure is essentially inverse to the Fock Poisson bracket on the set of shear coordinates. 

It is worth mentioning that a quantitative description of surfaces with marked points on the boundary has a long story: originated by Penner \cite{Penn1}, it was developed by Fock and Goncharov \cite{FG1}, Musiker, Schiffler and Williams \cite{MSW1}, \cite{MW}, and S. Fomin, M. Shapiro, and D. Thurston \cite{FST}, \cite{FT}. Note that the authors of \cite{FST} considered systems of (tagged) arcs starting and terminating either at bordered cusps or at punctures (holes) of $\Sigma_{g,s,n}$. In our approach, we allow only arcs starting and terminating at bordered cusps.

\section{Teichm\"uller spaces of Riemann surfaces with holes, and related $\lambda$-lengths (cluster variables)}\label{s:preliminaries}
\setcounter{equation}{0}

Following \cite{Penn1,Fock1,ChF2,ChSh,ChMaz2}, in this section we recall the combinatorial description of the Teichm\"uller space ${\mathfrak T}_{g,s,n}$ of Riemann surfaces of genus $g$ with $s>0$ holes/orbifold points, and with $n>0$ {\it bordered cusps} situated on the hole boundaries. These cusps are decorated by horocycles based at the endpoints of cusps at the absolute (and all ``geodesic distances from cusps'' are actually measured from crossings with these horocycles). 

\subsection{Fat graph description for Riemann surfaces $\Sigma_{g,s,n}$ and Teichm\"uller spaces}

\begin{definition}\label{def-pend}
We call a fat graph (a connected labelled graph with the prescribed cyclic ordering of edges entering each vertex) ${\mathcal G}_{g,s,n}$ a {\em spine of the Riemann surface} $\Sigma_{g,s,n}$ with $g$ handles, $s_h>0$ holes each containing $n_i>0$ bordered cusps ($\sum_{i} n_i=n>0$), and $s_o\ge 0$ holes/orbifold points without bordered cusps $(s=s_h+s_o)$ if
\begin{itemize}
\item[(a)] this graph can be embedded  without self-intersections in $\Sigma_{g,s,n}$;
\item[(b)] vertices of ${\mathcal G}_{g,s,n}$ are three-valent except exactly $n$ one-valent vertices incident to endpoints of $n$ {\it pending edges}; 
\item[(c)] every pending edge protrudes into a uniquely determined boundary component (a hole) and corresponds topologically to a marked point on the bounding perimeter of the corresponding hole, which therefore contains exactly $n_i>0$ marked points with a fixed cyclic ordering;
\item[(d)] upon cutting along all edges of ${\mathcal G}_{g,s,n}$ the Riemann surface $\Sigma_{g,s,n}$ splits into $s=s_h+s_o$ polygons each of which contains exactly one hole or an orbifold point and is simply connected upon erasing  this hole or orbifold point. All polygons containing orbifold points and holes without boundary cusps are monogons, every such monogon is bounded by an edge (a loop) that starts and terminates at the same three-valent vertex of the spine, and the number of loops is therefore exactly $s_o$;
\item[(e)] The edges in the above graph are labeled by distinct integers $\alpha,j,i=1,2,\dots,6g-6+3s+2n$, and we set a real number $Z_\alpha$ (a shear coordinate) into correspondence to the $\alpha$th edge if it is neither a pending edge, nor a loop. To each pending edge we set into correspondence a real number $\pi_j$ (an extended shear coordinate), and to each edge that
is a loop we set into correspondence the number $\omega_i$ (a coefficient) such that
$$
\omega_i=\left\{
\begin{array}{ll}
 2\cosh(P_i/2) & \hbox{if the monogon contains a hole with the perimeter $P_i\ge 0$},    \\
 2\cos(\pi/p_i) & \hbox{if the monogon contains an orbifold point of order  $p_i\in {\mathbb Z}_+$, $p_i\ge 2$}.
\end{array}
\right.
\label{omegai}
$$
\end{itemize}
\end{definition}

In the case of surfaces with bordered cusps, we have a preferred choice for endpoints of paths in a graph ${\mathcal G}_{g,s,n}$: the tips of cusps. We therefore expand the first homotopy group $\pi_1({\mathcal G}_{g,s,n})$ to a {\it groupoid of paths}: A \emph{path} $\mathfrak a_{j_1\to j_2}$ is a continuous line in ${\mathcal G}_{g,s,n}$ that starts and terminates at the corresponding one-valent vertices $j_1$ and $j_2$ (which may coincide) and has no backtrackings but is otherwise arbitrary. The groupoid property restricts the composition law $\mathfrak a_{j_1\to j_2} \circ \mathfrak a_{j_2\to j_3}=\mathfrak a_{j_1\to j_3}$, to only those paths that terminate and start at the same bordered cusp (in the resulting arc $\mathfrak a_{j_1\to j_3}$, we erase this intermediate endpoint together with  a backtracking unavoidably occurring in the composition process). 

Fat graphs ${\mathcal G}_{g,s,n}$ are in bijection with \emph{ideal triangle decompositions} of $\Sigma_{g,s,n}$  \cite{Penn1}: triangles correspond to three-valent vertices, vertices of ideal triangles are situated at the bordered cusps, whereas triangle sides, which are either inner sides shared by two distinct triangles, or two glued sides of the same triangle, or outer boundary sides, are in 1-1 correspondence with correspondingly inner edges, loops, or pending edges of the graph ${\mathcal G}_{g,s,n}$.

Paths $\mathfrak a_{i\to j}$ in ${\mathcal G}_{g,s,n}$ are in bijection with {\it arcs} on a Riemann surface $\Sigma_{g,s,n}$, which are directed geodesic lines that start and terminate at the bordered cusps and lie in the same homotopy classes as the corresponding paths. We fix the starting and terminating cusps of an arc to be right (w.r.t. the surface orientation) endpoints of boundary sides from the dual ideal triangle decomposition that are dual to the respective starting and terminating pending edges of the path. Note that all sides of ideal triangles of the ideal triangle decomposition dual to a given fat graph ${\mathcal G}_{g,s,n}$ are (arbitrarily directed) arcs themselves. Sides that are not paired are therefore boundary geodesics that connect neighboring bordered cusps on the same boundary component (a hole), and for these sides a preferable direction is the one compatible with the surface orientation.

A metrizable surface $\Sigma_{g,s,n}$ is the quotient of the Poincar\'e hyperbolic upper half plane under a discrete action of a Fuchsian group $\Delta_{g,s}\subset PSL(2,\mathbb R)$. Note that this group is insensitive to the presence of bordered cusps; to obtain a cusp structure on a boundary component we must explicitly draw geodesic lines (bounding arcs) located outside the hole perimeter and meeting each other pairwise at the absolute subsequently removing infinite domains separated by these bounding arcs from the bulk of the surface. We then obtain a ``crown-like'' (or ``diadem-like'' in the case of one cusp) structure put on top of perimeter lines of holes that contain bordered cusps. The standard statement in hyperbolic geometry is that conjugacy classes of hyperbolic elements of a Fuchsian group $\Delta_{g,s}$ are in the 1-1 correspondence with homotopy classes of closed paths in the Riemann surface $\Sigma_{g,s,n}$; explicitly for $\gamma\in \Delta_{g,s}$ we have that 
$$
\tr \gamma:=G_\gamma=e^{\ell_\gamma/2}+e^{-\ell_\gamma/2},
$$
where $\ell_\gamma$ is the length of a unique closed geodesic line belonging to this class and $G_\gamma$ is the corresponding geodesic function.

The real numbers $Z_\alpha$ in Definition~\ref{def-pend} are the {\em (Thurston) shear coordinates} (see \cite{ThSh},\cite{Bon2}). The real numbers $\pi_j$ are extended shear coordinates and we identify the set $\{Z_\alpha, \pi_j,w_i\}$ with coordinates of the decorated Teichm\"uller space ${\mathfrak T}_{g,s,n}$. It was proved in \cite{ChSh} that these sets parameterize all metrizable Riemann surfaces modulo a discretely acting \tcb{groupoid of flip morphisms} and vice versa, every such set corresponds to a metrizable Riemann surface. 

\begin{figure}[tb]
%\hspace*{2cm}
%\epsfysize=6cm
%\vskip .2in
{\psset{unit=1}
\begin{pspicture}(-2.5,-3)(2.5,2)
\newcommand{\FLIP}{%
{\psset{unit=1}
\psline[linewidth=18pt,linecolor=blue](0,-1)(0,1)
\psline[linewidth=18pt,linecolor=blue](0,1)(1.5,2)
\psline[linewidth=18pt,linecolor=blue](0,1)(-1.5,2)
\psline[linewidth=18pt,linecolor=blue](0,-1)(1.5,-2)
\psline[linewidth=18pt,linecolor=blue](0,-1)(-1.5,-2)
\psline[linewidth=14pt,linecolor=white](0,-1)(0,1)
\psline[linewidth=14pt,linecolor=white](0,1)(1.5,2)
\psline[linewidth=14pt,linecolor=white](0,1)(-1.5,2)
\psline[linewidth=14pt,linecolor=white](0,-1)(1.5,-2)
\psline[linewidth=14pt,linecolor=white](0,-1)(-1.5,-2)
\psbezier[linecolor=red, linestyle=dashed, linewidth=1.5pt](1.8,0)(1.1,0)(0,-1)(0,-2)
\psbezier[linecolor=red, linestyle=dashed, linewidth=1.5pt](-1.8,0)(-1.1,0)(0,-1)(0,-2)
\psbezier[linecolor=red, linestyle=dashed, linewidth=1.5pt](1.8,0)(1.1,0)(0,1)(0,2)
\psbezier[linecolor=red, linestyle=dashed, linewidth=1.5pt](-1.8,0)(-1.1,0)(0,1)(0,2)
\psline[linecolor=red, linestyle=dashed, linewidth=1.5pt](1.8,0)(-1.8,0)
}
}
\rput(0,0){\FLIP}
\rput(-1.2,0.6){\makebox(0,0)[cb]{$\lambda_a$}}
\rput(1.2,0.6){\makebox(0,0)[cb]{$\lambda_b$}}
\rput(-1.2,-0.6){\makebox(0,0)[ct]{$\lambda_d$}}
\rput(1.2,-0.6){\makebox(0,0)[ct]{$\lambda_c$}}
\rput(0.5,.3){\makebox(0,0){$Z_e$}}
\end{pspicture}
\begin{pspicture}(-2.5,-3)(2.5,2)
\newcommand{\FLIP}{%
{\psset{unit=1}
\psline[linewidth=18pt,linecolor=blue](0,-0.5)(0,1)
\psline[linewidth=18pt,linecolor=blue](0,1)(1.5,2)
\psline[linewidth=18pt,linecolor=blue](0,1)(-1.5,2)
%\psline[linewidth=18pt,linecolor=blue](0,-1)(1.5,-2)
%\psline[linewidth=18pt,linecolor=blue](0,-1)(-1.5,-2)
%
\psline[linewidth=14pt,linecolor=white](0,-0.6)(0,1)
\psline[linewidth=14pt,linecolor=white](0,1)(1.5,2)
\psline[linewidth=14pt,linecolor=white](0,1)(-1.5,2)
%\psline[linewidth=14pt,linecolor=white](0,-1)(1.5,-2)
%\psline[linewidth=14pt,linecolor=white](0,-1)(-1.5,-2)
%\psbezier[linecolor=red, linestyle=dashed, linewidth=1.5pt](1.8,0)(1.1,0)(0,-1)(0,-2)
%\psbezier[linecolor=red, linestyle=dashed, linewidth=1.5pt](-1.8,0)(-1.1,0)(0,-1)(0,-2)
\psbezier[linecolor=red, linestyle=dashed, linewidth=1.5pt](1.8,0)(1.1,0)(0,1)(0,2)
\psbezier[linecolor=red, linestyle=dashed, linewidth=1.5pt](-1.8,0)(-1.1,0)(0,1)(0,2)
\psline[linecolor=red, linestyle=dashed, linewidth=1.5pt](1.8,0)(-1.8,0)
}
}
\rput(0,0){\FLIP}
\rput(-1.2,0.6){\makebox(0,0)[cb]{$\lambda_a$}}
\rput(1.2,0.6){\makebox(0,0)[cb]{$\lambda_b$}}
\rput(-0.5,-0.2){\makebox(0,0)[rt]{$\lambda_c$}}
\rput(0.5,.3){\makebox(0,0){$\pi_c$}}
\end{pspicture}
}
\caption{\small Correspondence between (extended) shear coordinates $Z,\pi$ and $\lambda$-lengths.}
\label{fi:dual}
\end{figure}
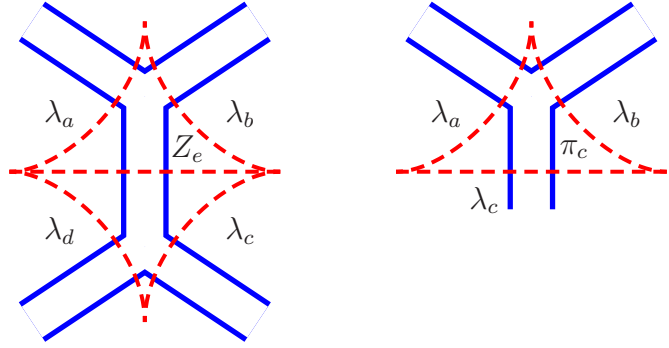

We set into correspondence to every arc its $\lambda$-length, $\lambda_{\mathfrak a} =e^{l_{\mathfrak a}/2}$---the exponential of a half of the signed length of a part of a geodesic arc $\mathfrak a$ stretched between the corresponding decorated cusps (the sign is negative if the horocycles decorating these cusps intersect). The extended shear coordinates are related to the $\lambda$-lengths through the cross-ratio relation (see Fig.~\ref{fi:dual})
\be
e^{Z_e}=\frac{\lambda_a\lambda_c}{\lambda_b\lambda_d},\qquad e^{\pi_c}=\frac{\lambda_a\lambda_b}{\lambda_c},
\label{cross-l}
\ee
where we label by the same index both the fat graph edge that is not a loop and the incident to this edge arc from the ideal triangle decomposition dual to the given graph.  The geometrical meaning of $Z_e$ is the signed geodesic distance between perpendiculars to the common side $e$ of two adjacent ideal triangles through the vertices of these triangles.

\subsection{The Fuchsian group $\Delta_{g,s}$ and geodesic functions}\label{ss:geodesic}

The combinatorial description of the groupoid of paths, or, equivalently, elements of the {\em decorated character variety} $(SL(2,\mathbb R))^{2g+s+n-2}/U^n$ (see \cite{CMR2}) is given in terms of edge and turn matrices.  Every time the path in a graph ${\mathcal G}_{g,s,n}$ goes along $\alpha$th inner edge or starts or terminate at a pending edge, we insert~\cite{Fock1} the so-called {\it edge matrix}  $X_{Z_\alpha}$ or $ X_{\pi_j}$,
\be
\label{XZ} X_{Y}=\left(
\begin{array}{cc} 0 & -e^{Y/2}\\
                e^{-Y/2} & 0\end{array}\right),\quad Y=Z_\alpha\ \hbox{or}\ Y=\pi_j,
\ee
in the proper place into the corresponding product of matrices (all matrix multiplications are from right to left). When a path makes right or left turn at a three-valent vertex, we insert the corresponding ``right'' and ``left'' turn matrices
\be
\label{R}
R=\left(\begin{array}{cc} 1 & 1\\ -1 & 0\end{array}\right), \qquad
L= R^2=\left(\begin{array}{cc} 0 & 1\\ -1 &
-1\end{array}\right),
\ee
and finally, when a path is going along an $i$th loop clockwise(counterclockwise), we insert the matrix $F_{\omega_i}$ (or $-F^{-1}_{\omega_i}$):
\be
\label{F-p}
F_{\omega_i}:=
\left(\begin{array}{cc} 0 & 1\\ -1 & -w_i\end{array}\right),\qquad -F^{-1}_{\omega_i}:=
\left(\begin{array}{cc} w_i & 1\\ -1 & 0\end{array}\right).
\ee

An element $P_{\mathfrak a}\in PSL(2,\mathbb R)$ in the groupoid of paths has then the typical structure:
\be
\label{Pgamma}
P_{\mathfrak a}=X_{\pi_{j_2}}LX_{Z_n}RX_{Z_{n-1}}\cdots RX_{Z_{k+1}}LX_{Z_k}F_{\omega_i} X_{Z_{k-1}}R\dots X_{Z_1}L X_{\pi_{j_1}}
\ee
for a path starting at cusp $j_1$ and terminating at cusp $j_2$. The $\lambda$-length of this path is then the {\em upper-right matrix element} of $P_{\mathfrak a}$ (and it is easy to see that it does not depend on the path direction). 

We obtain a realization of the Fuchsian group if we choose one of the bordered cusps (say, $j$th cusp) and consider a group of all paths starting and terminating at this cusp. Geodesic functions are then given by traces of the corresponding elements 
\be
\label{G}
G_{\gamma}\equiv \tr P_{\mathfrak a_{jj}}=2\cosh(\ell_\gamma/2),
\ee
where $\ell_\gamma$ is the length of the closed geodesic that is homeomorphic to the arc $\mathfrak a_{jj}$ upon erasing the endpoint.

This construction implies the following fundamental property: For any graph $\mathcal G_{g,s,n}$ all matrix elements of $P_{\mathfrak a}$ for every arc ${\mathfrak a}$ (and, correspondingly, all $\lambda$-lengths and geodesic functions $G_\gamma$ are polynomials with sign-definite integer coefficients: $\lambda_{\mathfrak a}\in \mathbb Z_+[[e^{\pi_j/2}, e^{\pm Z_\alpha/2}, \omega_i]]$ and $G_{\gamma}\in \mathbb Z_+[[e^{\pm Z_\alpha/2}, \omega_i]]$.

If we have at least one bordered cusp, we can invert formulas (\ref{cross-l}) and express $\lambda$-lengths in terms of extended shear coordinates. The polynomial expressions obtained from (\ref{Pgamma}) for $\lambda$-lengths of arcs become monoidal for arcs constituting the ideal triangle decomposition dual to a given fat graph  ${\mathcal G}_{g,s,n}$.

\begin{lemma}\label{lem:lambda}\cite{ChMaz2}
Paths $P_{\mathfrak a_\alpha}$ corresponding to arcs $\mathfrak a_\alpha$ from the ideal triangle decomposition dual to a given fat graph  ${\mathcal G}_{g,s,n}$ and labelled by edges $\alpha,j$ (that are not loops, and the first and the last edge are pending edges) of this graph have the product structure  $P_{\mathfrak a_\alpha}=X_{\pi_2} R X_{Z_{\beta_n}}\cdot R\cdots R F_w R\cdots R X_{Z_\alpha} L X_{Z_{\beta_j}} \cdots X_{Z_{\beta_1}} L X_{\pi_1}$, and their upper-right elements are merely
$$
e^{l_{\mathfrak a_\alpha}/2}=e^{(\pi_1+\pi_2+Z_\alpha +\sum Z_\beta)/2},
$$
where the sum ranges (with multiplicities) all edges along the path $\mathfrak a_\alpha$. Note that coefficients $w_i$ do not contribute to these $\lambda$-lengths.
\end{lemma}

\subsection{Poisson and symplectic structures}\label{ss:Poisson}

One of the most attractive properties of the graph description is a very simple Poisson algebra on the set of coordinates $Z_\alpha$, $\pi_j$ due to V.V. Fock \cite{F97}: let $Y_k$, $k=1,2,3 \mod 3$, denote either $Z$- or $\pi$-variables of cyclically ordered edges incident to a three-valent vertex. The Poisson bi-vector field is then
\be
\label{WP-PB}
w_{\text{WP}}:=\sum_{{\hbox{\small 3-valent} \atop \hbox{\small vertices} }}
\,\sum_{k=1}^{3} \partial_{Y_k}\wedge \partial_{Y_{k+1}},
\ee

\begin{theorem}\label{th-WP}\cite{ChF2} 
The bracket \eqref{WP-PB} gives rise to the {\em Goldman
bracket}  \cite{Gold} on the space of geodesic functions.
\end{theorem}

The center of the Poisson algebra {\rm(\ref{WP-PB})} is generated by sums of $Z_\alpha$ and $\pi_j$ (taken with multiplicities) of all edges incident to a given hole, so together with the coefficients $w_i$ we have exactly $s$ central elements.

At the same time, we have a strikingly similar expression for the {\em symplectic structure} on the set of $\lambda$-lengths introduced by Penner \cite{Penn92}:
\be
\label{WP-SS}
\Omega_{\text{WP}}:=\sum_{{\hbox{\small ideal} \atop \hbox{\small triangles} }}
\,\sum_{k=1}^{3}d\log \lambda_k\wedge d\log \lambda_{k+1}.
\ee
Despite the simplicity of the above formulas, before introducing bordered cusps all attempts of constructing Poisson brackets for $\lambda$ lengths or symplectic structure for shear coordinates were marred with discrepancies. But now, armed with explicit formulas from Lemma~\ref{lem:lambda}, we can solve this problem. First, in \cite{ChMaz2} the author together with Marta Mazzocco found Poisson and quantum structures on the set of $\lambda$-lengths; the result reproduced Berenstein--Zelevinsky quantum cluster algebras \cite{BerZel}. The last missing item was a symplectic form on the set of extended shear coordinates; we complete it in Theorem~\ref{th:Omega} in the next section. 

\subsection{Flip morphisms of fat graphs}\label{ss:flip}

Transitions between different parameterizations are formulated in terms of mapping class group transformations obtained by composing flip morphisms
(mutations) of edges: any two spines from the given topological class are related by a finite sequence of flips.

There are two sorts of flip morphisms: those induced by flips of inner edges (see Fig. \ref{fi:flip}) and those induced by flips of edges that are adjacent to a loop (see Fig. \ref{fi:interchange-p-dual}); no flips can be performed on pending edges.

\subsubsection{Flipping inner edges}\label{sss:mcg}

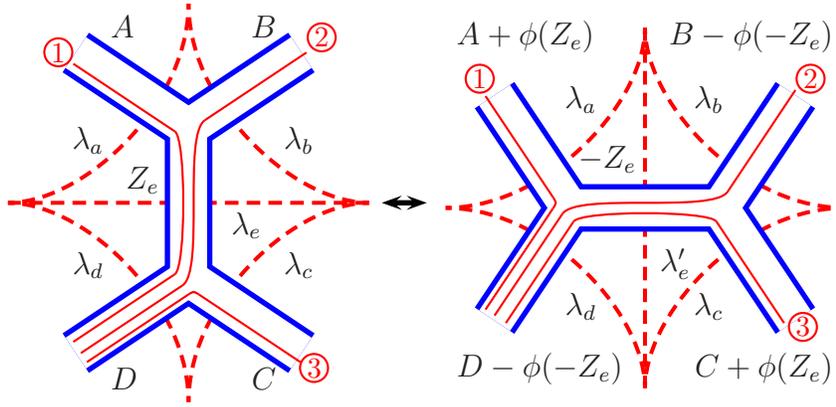
\begin{figure}[tb]
\begin{pspicture}(-3,-3)(4,3){
\newcommand{\FLIP}{%
{\psset{unit=1}
\psline[linewidth=18pt,linecolor=blue](0,-1)(0,1)
\psline[linewidth=18pt,linecolor=blue](0,1)(1.5,2)
\psline[linewidth=18pt,linecolor=blue](0,1)(-1.5,2)
\psline[linewidth=18pt,linecolor=blue](0,-1)(1.5,-2)
\psline[linewidth=18pt,linecolor=blue](0,-1)(-1.5,-2)
\psline[linewidth=14pt,linecolor=white](0,-1)(0,1)
\psline[linewidth=14pt,linecolor=white](0,1)(1.5,2)
\psline[linewidth=14pt,linecolor=white](0,1)(-1.5,2)
\psline[linewidth=14pt,linecolor=white](0,-1)(1.5,-2)
\psline[linewidth=14pt,linecolor=white](0,-1)(-1.5,-2)
}
}
\rput(-2.5,0){\psbezier[linecolor=red, linestyle=dashed, linewidth=1.5pt](2.4,0)(1.4,0)(0,-1.3)(0,-2.65)
\psbezier[linecolor=red, linestyle=dashed, linewidth=1.5pt](-2.4,0)(-1.4,0)(0,-1.3)(0,-2.65)
\psbezier[linecolor=red, linestyle=dashed, linewidth=1.5pt](2.4,0)(1.4,0)(0,1.3)(0,2.65)
\psbezier[linecolor=red, linestyle=dashed, linewidth=1.5pt](-2.4,0)(-1.4,0)(0,1.3)(0,2.65)
\psline[linecolor=red, linestyle=dashed, linewidth=1.5pt](2.4,0)(-2.4,0)
\FLIP}
\rput(.3,0){\psline[linewidth=2pt]{<->}(-0.3,0)(0.3,0)}
\rput{90}(3.5,0){\psbezier[linecolor=red, linestyle=dashed, linewidth=1.5pt](2.4,0)(1.4,0)(0,-1.3)(0,-2.65)
\psbezier[linecolor=red, linestyle=dashed, linewidth=1.5pt](-2.4,0)(-1.4,0)(0,-1.3)(0,-2.65)
\psbezier[linecolor=red, linestyle=dashed, linewidth=1.5pt](2.4,0)(1.4,0)(0,1.3)(0,2.65)
\psbezier[linecolor=red, linestyle=dashed, linewidth=1.5pt](-2.4,0)(-1.4,0)(0,1.3)(0,2.65)
\psline[linecolor=red, linestyle=dashed, linewidth=1.5pt](2.4,0)(-2.4,0)
\FLIP}
\rput(-2.5,0){
\rput(-1.1,2.2){\makebox(0,0)[lb]{$A$}}
\rput(1.1,2.2){\makebox(0,0)[rb]{$B$}}
\rput(-0.9,0.3){\makebox(0,0)[lc]{$Z_e$}}
\rput(1.1,-2.2){\makebox(0,0)[rt]{$C$}}
\rput(-1.1,-2.2){\makebox(0,0)[lt]{$D$}}
\rput(0.9,-0.3){\makebox(0,0)[rc]{$\lambda_e$}}
\rput(-1.2,0.65){\makebox(0,0)[rb]{$\lambda_a$}}
\rput(1.2,0.65){\makebox(0,0)[lb]{$\lambda_b$}}
\rput(1.2,-0.65){\makebox(0,0)[lt]{$\lambda_c$}}
\rput(-1.2,-0.65){\makebox(0,0)[rt]{$\lambda_d$}}
\psline[linecolor=red](-1.5,-2)(-.15,-1.1)
\psline[linecolor=red](1.5,2)(.15,1.1)
\psbezier[linecolor=red](-.15,-1.1)(0.15,-0.9)(-0.15,0.9)(.15,1.1)
\rput(-.1,.15){\psline[linecolor=red](-1.5,-2)(-.15,-1.1)}
\rput(-.1,-.15){\psline[linecolor=red](-1.5,2)(-.15,1.1)}
\psbezier[linecolor=red](-.25,-.95)(-0.1,-0.85)(-0.1,0.85)(-.25,.95)
\rput(.08,-.12){\psline[linecolor=red](-1.5,-2)(-.15,-1.1)}
\rput(-.08,-.12){\psline[linecolor=red](1.5,-2)(.15,-1.1)}
\psbezier[linecolor=red](-.07,-1.22)(0,-1.176)(0,-1.176)(.07,-1.22)
}
\rput(3.5,0){
\rput(-2.5,-2){\makebox(0,0)[lt]{$D - \phi(-Z_e)$}}
\rput(2.5,-2){\makebox(0,0)[rt]{$C+\phi(Z_e)$}}
\rput(2.5,2){\makebox(0,0)[rb]{$B-\phi(-Z_e)$}}
\rput(-2.5,2){\makebox(0,0)[lb]{$A+\phi(Z_e)$}}
\rput(-0.5,0.4){\makebox(0,0)[cb]{$-Z_e$}}
\rput(0.4,-0.6){\makebox(0,0)[ct]{$\lambda'_{e}$}}
\rput(-0.65,1.2){\makebox(0,0)[rb]{$\lambda_a$}}
\rput(0.65,1.2){\makebox(0,0)[lb]{$\lambda_b$}}
\rput(0.65,-1.2){\makebox(0,0)[lt]{$\lambda_c$}}
\rput(-0.65,-1.2){\makebox(0,0)[rt]{$\lambda_d$}}
}
\rput{90}(3.5,0){
\psline[linecolor=red](1.5,-2)(.15,-1.1)
\psline[linecolor=red](-1.5,2)(-.15,1.1)
\psbezier[linecolor=red](.15,-1.1)(-0.15,-0.9)(0.15,0.9)(-.15,1.1)
\rput(-.1,.15){\psline[linecolor=red](-1.5,-2)(-.15,-1.1)}
\rput(-.1,-.15){\psline[linecolor=red](-1.5,2)(-.15,1.1)}
\psbezier[linecolor=red](-.25,-.95)(-0.1,-0.85)(-0.1,0.85)(-.25,.95)
}
\rput{270}(3.5,0){
\rput(.08,-.12){\psline[linecolor=red](-1.5,-2)(-.15,-1.1)}
\rput(-.08,-.12){\psline[linecolor=red](1.5,-2)(.15,-1.1)}
\psbezier[linecolor=red](-.07,-1.22)(0,-1.176)(0,-1.176)(.07,-1.22)
}
% added geodesic lines
\rput(-2.5,0){
\put(-1.8,1.9){\makebox(0,0)[cc]{\hbox{\tcr{\small$1$}}}}
\put(-1.8,1.9){\pscircle[linecolor=red]{.2}}
\put(1.7,2.1){\makebox(0,0)[cc]{\hbox{\tcr{\small$2$}}}}
\put(1.7,2.1){\pscircle[linecolor=red]{.2}}
\put(1.6,-2.3){\makebox(0,0)[cc]{\hbox{\tcr{\small$3$}}}}
\put(1.6,-2.3){\pscircle[linecolor=red]{.2}}
}
%right half
\rput(3.5,0){
\put(-2.2,1.7){\makebox(0,0)[cc]{\hbox{\tcr{\small$1$}}}}
\put(-2.2,1.7){\pscircle[linecolor=red]{.2}}
\put(2.2,1.7){\makebox(0,0)[cc]{\hbox{\tcr{\small$2$}}}}
\put(2.2,1.7){\pscircle[linecolor=red]{.2}}
\put(2.1,-1.6){\makebox(0,0)[cc]{\hbox{\tcr{\small$3$}}}}
\put(2.1,-1.6){\pscircle[linecolor=red]{.2}}
}
}
\end{pspicture}
\caption{\small Flip on an internal edge (labeled ``$e$'') that is neither a loop nor adjacent to a loop. We indicate the correspondences between geodesic paths undergoing the flip. Dashed lines are edges of the dual ideal triangle decomposition.}
\label{fi:flip}
\end{figure}

\begin{lemma} \label{lem-abc}~\cite{ChF1,ChF2}
In the notation of Fig.~\ref{fi:flip}, the transformation
$$
({\tilde A},{\tilde B},{\tilde C},{\tilde D},{\tilde Z_e})=(A+\phi(Z_e), B-\phi(-Z_e), C+\phi(Z_e), D-\phi(-Z_e), -Z_e)
%\label{abc}
$$
where $\phi (Z)={\rm log}(1+e^Z)$,
preserves the paths {\rm(\ref{G})} (thus preserving both geodesic functions and $\lambda$-lengths)  simultaneously preserving
Poisson structure {\rm(\ref{WP-PB})} on the shear coordinates. The dual Ptolemy transformation of $\lambda$-lengths (cluster mutation),
$\lambda_e\lambda'_{e}=\lambda_a\lambda_c+\lambda_b\lambda_d$ preserves the symplectic structure (\ref{WP-SS}).
\end{lemma}

The proof is local w.r.t. the graph $\mathcal G_{g,s,n}$ and is based on matrix equalities
\bea
X_DRX_{Z_e}RX_{A}&=&X_{\tilde A}RX_{\tilde D},\nonumber\\
X_DRX_{Z_e}LX_B&=&X_{\tilde D}LX_{\tilde Z_e}RX_{\tilde B},\nonumber\\
X_CLX_D&=&X_{\tilde C}LX_{\tilde Z_e}LX_{\tilde D},\nonumber
\eea
each pertaining to the corresponding geodesic in Fig.~\ref{fi:flip}).

\subsubsection{Flipping the edge incident to a loop}\label{sss:pending}

\begin{figure}[tb]
\begin{pspicture}(-3,-2.5)(4,2.5){
\newcommand{\FLIP}{%
{\psset{unit=1}
\psbezier[linewidth=1.5pt, linestyle=dashed, linecolor=red](-5,0)(-3,2.5)(0,2.5)(2,0)
\psbezier[linewidth=1.5pt, linestyle=dashed, linecolor=red](-5,0)(-3,-2.5)(0,-2.5)(2,0)
\psbezier[linewidth=1.5pt, linestyle=dashed, linecolor=red](-5,0)(-3.5,1.6)(-1,1.6)(-1,0)
\psbezier[linewidth=1.5pt, linestyle=dashed, linecolor=red](-5,0)(-3.5,-1.6)(-1,-1.6)(-1,0)
\psline[linewidth=18pt,linecolor=blue](-2,0)(0,0)
\psline[linewidth=18pt,linecolor=blue](0,0)(1,1.5)
\psline[linewidth=18pt,linecolor=blue](0,0)(1,-1.5)
\pscircle[linewidth=2pt,linecolor=blue,fillstyle=solid,fillcolor=white](-2.8,0){1}
\psline[linewidth=14pt,linecolor=white](-2,0)(0,0)
\psline[linewidth=14pt,linecolor=white](0,0)(1,1.5)
\psline[linewidth=14pt,linecolor=white](0,0)(1,-1.5)
\rput(-2.8,0){\pscircle[linecolor=blue,linewidth=2pt,fillstyle=solid,fillcolor=white](0,0){0.5}}
\psline[linewidth=1.5pt, linestyle=dashed, linecolor=red](-5,0)(-3.8,0)
\psline[linewidth=1.5pt, linestyle=dashed, linecolor=red](-2.8,0)(-3.3,0)
%\rput(-2,0){\pscircle*{0.05}}
\psarc[linecolor=red](-2,0.3){0.1}{200}{270}
\psarc[linecolor=red](-2,0.3){0.2}{200}{270}
\psarc[linecolor=red](-2,-0.3){0.1}{90}{160}
\psarc[linecolor=red](-2,-0.3){0.2}{90}{160}
\psline[linecolor=red](-2,0.1)(-0.01,0.1)
\psline[linecolor=red](-2,0.2)(-0,0.2)
\psline[linecolor=red](-2,-0.1)(-0.01,-0.1)
\psline[linecolor=red](-2,-0.2)(-0,-0.2)
\psarc[linecolor=red](-2.8,0){0.75}{20}{340}
\psarc[linecolor=red](-2.8,0){0.65}{20}{340}
}
}
\rput(-2.5,0){\FLIP}
\psline[linewidth=2pt]{<->}(-0.3,0)(0.3,0)
\rput{180}(2.5,0){\FLIP}
\rput(-2.5,0){
\rput(0.2,1.5){\makebox(0,0)[lb]{$A$}}
\rput(-0.8,0.5){\makebox(0,0)[lb]{$Z_e$}}
\rput(0.2,-1.5){\makebox(0,0)[lt]{$B$}}
\rput(1.3,0.8){\makebox(0,0)[lb]{$\lambda_a$}}
\rput(1.3,-0.8){\makebox(0,0)[lt]{$\lambda_b$}}
\rput(-1,-0.5){\makebox(0,0)[lt]{$\lambda_e$}}
\rput(-1.2,2){\makebox(0,0)[rb]{$\displaystyle w=e^\xi+e^{-\xi},\ \xi\in\mathbb C$}}
\rput(-4,0.2){\makebox(0,0)[rb]{$\omega$}}
}
\rput(3.5,0){
\rput(-2.3,0.8){\makebox(0,0)[rb]{$\lambda_a$}}
\rput(-2.3,-0.8){\makebox(0,0)[rt]{$\lambda_b$}}
\rput(0,-0.5){\makebox(0,0)[rt]{$\lambda'_e$}}
\rput(-1.4,-2){\makebox(0,0)[lt]{$\displaystyle B - \phi(-Z_e+\xi)-\phi(-Z_e-\xi)$}}
\rput(-1.4,2){\makebox(0,0)[lb]{$\displaystyle A + \phi(Z_e+\xi)+\phi(Z_e-\xi)$}}
\rput(-0.1,0.5){\makebox(0,0)[rb]{$-Z_e$}}
\rput(3,0.2){\makebox(0,0)[lb]{$\omega$}}
}
\rput{90}(3.5,0){
%\rput(-.1,.15){\psline[linecolor=red](-1.5,-2)(-.15,-1.1)}
\rput(-.1,-.15){\psline[linecolor=red](-1.5,2)(-.09,1.06)}
\rput(.1,-.15){\psline[linecolor=red](1.5,2)(.09,1.06)}
%\psbezier[linecolor=red](-.25,-.95)(-0.1,-0.85)(-0.1,0.85)(-.25,.95)
}
\rput{270}(-3.5,0){
%\rput(-.1,.15){\psline[linecolor=red](-1.5,-2)(-.15,-1.1)}
\rput(-.1,-.15){\psline[linecolor=red](-1.5,2)(-.09,1.06)}
\rput(.1,-.15){\psline[linecolor=red](1.5,2)(.09,1.06)}
}
\rput{270}(3.5,0){
\rput(.08,-.12){\psline[linecolor=red](-1.5,-2)(-.15,-1.1)}
\rput(-.08,-.12){\psline[linecolor=red](1.5,-2)(.15,-1.1)}
\psbezier[linecolor=red](-.07,-1.22)(0,-1.176)(0,-1.176)(.07,-1.22)
}
\rput{90}(-3.5,0){
\rput(.08,-.12){\psline[linecolor=red](-1.5,-2)(-.15,-1.1)}
\rput(-.08,-.12){\psline[linecolor=red](1.5,-2)(.15,-1.1)}
\psbezier[linecolor=red](-.07,-1.22)(0,-1.176)(0,-1.176)(.07,-1.22)
}
\rput(-2.5,0){
\rput(-.06,.04){\psline[linecolor=red](1,1.5)(0.1,.15)}
\rput(.06,-.04){\psline[linecolor=red](1,1.5)(0.1,.15)}
\psbezier[linecolor=red](.04,.19)(-0.02,0.1)(-0.02,0.1)(-0.1,0.1)
\psbezier[linecolor=red](.16,.11)(0.02,-0.1)(0.02,-.1)(-0.1,-.1)
}
\rput(2.5,0){
\rput(.06,.04){\psline[linecolor=red](-1,1.5)(-0.1,.15)}
\rput(-.06,-.04){\psline[linecolor=red](-1,1.5)(-0.1,.15)}
\psbezier[linecolor=red](-.04,.19)(0.02,0.1)(0.02,0.1)(0.1,0.1)
\psbezier[linecolor=red](-.16,.11)(-0.02,-0.1)(-0.02,-.1)(0.1,-.1)
}
% added geodesic lines
\rput(-3.5,0){
\put(2.3,-1.6){\makebox(0,0)[cc]{\hbox{\tcr{\small$3$}}}}
\put(2.3,-1.6){\pscircle[linecolor=red]{.2}}
\put(2.2,1.7){\makebox(0,0)[cc]{\hbox{\tcr{\small$2$}}}}
\put(2.2,1.7){\pscircle[linecolor=red]{.2}}
\put(1.95,-1.8){\makebox(0,0)[cc]{\hbox{\tcr{\small$1$}}}}
\put(1.95,-1.8){\pscircle[linecolor=red]{.2}}
}
%right half
\rput(3.5,0){
\put(-2.3,-1.6){\makebox(0,0)[cc]{\hbox{\tcr{\small$1$}}}}
\put(-2.3,-1.6){\pscircle[linecolor=red]{.2}}
\put(-2.2,1.7){\makebox(0,0)[cc]{\hbox{\tcr{\small$2$}}}}
\put(-2.2,1.7){\pscircle[linecolor=red]{.2}}
\put(-1.95,-1.8){\makebox(0,0)[cc]{\hbox{\tcr{\small$3$}}}}
\put(-1.95,-1.8){\pscircle[linecolor=red]{.2}}
}
}
\end{pspicture}
\caption{\small
The transformation of {shear coordinates} when flipping an edge incident to a loop;
$w=e^\xi+e^{-\xi}$ with $\xi\in\mathbb C$. We indicate how geodesic lines change upon flipping the edge. Dashed lines are edges of the dual ideal triangle decomposition.
}
\label{fi:interchange-p-dual}
\end{figure}
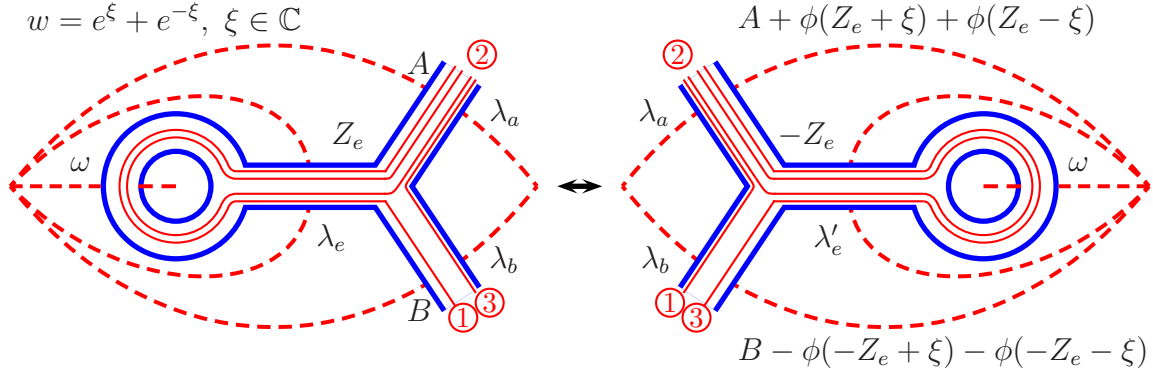

\begin{lemma} \label{lem-pending1}{\rm (\cite{ChSh},\cite{ChMaz2})}
The transformation~in Fig.~\ref{fi:interchange-p-dual}
$$
\{\tilde A,\tilde B,\tilde Z_e\}:= \{A+\phi(Z_e+\xi)+\phi(Z_e-\xi), B-\phi(-Z_e+\xi)-\phi(-Z_e-\xi),-Z_e\},\quad w=e^\xi+e^{-\xi},
$$
where $\phi(x)=\log(1+e^x)$ and $\xi\in\mathbb C$ is a morphism of the space ${\mathfrak T}_{g,s,n}$ that  preserves both Poisson structures {\rm(\ref{WP-PB})} and the path functions. The dual transformation (generalized cluster transformation) $\lambda_e\lambda'_{e}=\lambda_a^2+w\lambda_a\lambda_b+\lambda_b^2$ preserves the symplectic structure  (\ref{WP-SS}).
\end{lemma}

\section{Symplectic structure for extended shear coordinates}

%We have the following theorem.

\begin{theorem}\label{th:Omega}
The symplectic structure $\Omega_{\text{WP}}$ expressed in extended shear coordinates $\{Z_\alpha,\pi_j\}(=Y_\beta)$ on $\mathfrak T_{g,s,n}$ is 
\be\label{star}
\Omega_{\text{WP}}=\sum_{j=1}^n \Omega_j,\quad \Omega_j=\sum_{\beta_i<\beta_2,\ \beta\in I_j} dY_{\beta_1}\wedge dY_{\beta_2},
\ee
where the sum w.r.t. $j$ ranges all {\em windows}---parts of boundary components confined between boundary cusps (pending edges)---and $I_j$ are sets of indices of edges (possibly with repetitions) incident to the $j$th window with a natural linear ordering of edges compatible with the surface orientation. (This ordering is equivalent to introducing a ciliation in \cite{Fock-Rosly}.)
\end{theorem}

\begin{figure}[tb]
%\hspace*{2cm}
%\epsfysize=6cm
%\vskip .2in
{\psset{unit=1}
\begin{pspicture}(-2.5,-3)(2.5,2)
\newcommand{\BRANCHBLUE}{%
{\psset{unit=1}
\psline[linewidth=18pt,linecolor=blue](0,0.2)(0,1.2)
\psline[linewidth=18pt,linecolor=blue](0,1.2)(-.8,1.6)
\psline[linewidth=18pt,linecolor=blue](0,1.2)(.8,1.6)
\psline[linewidth=18pt,linecolor=blue](0.8,1.6)(1.4,1.4)
\psline[linewidth=18pt,linecolor=blue](0.8,1.6)(0.8,2.2)
\psline[linewidth=18pt,linecolor=blue](-0.8,1.6)(-1.4,1.4)
\psline[linewidth=18pt,linecolor=blue](-0.8,1.6)(-0.8,2.2)
\psline[linewidth=14pt,linecolor=white](0,0)(0,1.2)
\psline[linewidth=14pt,linecolor=white](0,1.2)(-.8,1.6)
\psline[linewidth=14pt,linecolor=white](0,1.2)(.8,1.6)
\psline[linewidth=14pt,linecolor=white](0.8,1.6)(1.4,1.4)
\psline[linewidth=14pt,linecolor=white](0.8,1.6)(0.8,2.2)
\psline[linewidth=14pt,linecolor=white](-0.8,1.6)(-1.4,1.4)
\psline[linewidth=14pt,linecolor=white](-0.8,1.6)(-0.8,2.2)
%
%\psline[linewidth=14pt,linecolor=white](0,-1)(0,1)
%\psline[linewidth=14pt,linecolor=white](0,1)(1.5,2)
%\psline[linewidth=14pt,linecolor=white](0,1)(-1.5,2)
%\psline[linewidth=14pt,linecolor=white](0,-1)(1.5,-2)
%\psline[linewidth=14pt,linecolor=white](0,-1)(-1.5,-2)
%\psbezier[linecolor=red, linestyle=dashed, linewidth=1.5pt](1.8,0)(1.1,0)(0,-1)(0,-2)
%\psbezier[linecolor=red, linestyle=dashed, linewidth=1.5pt](-1.8,0)(-1.1,0)(0,-1)(0,-2)
%\psbezier[linecolor=red, linestyle=dashed, linewidth=1.5pt](1.8,0)(1.1,0)(0,1)(0,2)
%\psbezier[linecolor=red, linestyle=dashed, linewidth=1.5pt](-1.8,0)(-1.1,0)(0,1)(0,2)
%\psline[linecolor=red, linestyle=dashed, linewidth=1.5pt](1.8,0)(-1.8,0)
}
}
\rput(0,0){\BRANCHBLUE}
\rput{120}(-0.1,0.1){\BRANCHBLUE}
\rput{-120}(-0.1,-0){\BRANCHBLUE}
\rput(0,0){
\psline[linewidth=1pt,linecolor=red](0,0.2)(0,1.2)
\psline[linewidth=1pt,linecolor=red](0,1.2)(.8,1.6)
\psline[linewidth=1pt,linecolor=red](0.8,1.6)(1.4,1.4)
\psline[linewidth=1pt,linecolor=green](0.1,-0.15)(0.1,1.1)
\psline[linewidth=1pt,linecolor=green](0.1,1.1)(.75,1.45)
\psline[linewidth=1pt,linecolor=green](0.75,1.45)(1.4,1.2)}
\rput{120}(-0.1,0.1){
\psline[linewidth=1pt,linecolor=magenta](0,0.2)(0,1.2)
\psline[linewidth=1pt,linecolor=magenta](0,1.2)(.8,1.6)
\psline[linewidth=1pt,linecolor=magenta](0.8,1.6)(1.4,1.4)
\psline[linewidth=1pt,linecolor=red](0.1,-0.15)(0.1,1.1)
\psline[linewidth=1pt,linecolor=red](0.1,1.1)(.75,1.45)
\psline[linewidth=1pt,linecolor=red](0.75,1.45)(1.4,1.2)}
\rput{-120}(-0.1,-0){
\psline[linewidth=1pt,linecolor=green](0,0.2)(0,1.2)
\psline[linewidth=1pt,linecolor=green](0,1.2)(.8,1.6)
\psline[linewidth=1pt,linecolor=green](0.8,1.6)(1.4,1.4)
\psline[linewidth=1pt,linecolor=magenta](0.1,-0.15)(0.1,1.1)
\psline[linewidth=1pt,linecolor=magenta](0.1,1.1)(.75,1.45)
\psline[linewidth=1pt,linecolor=magenta](0.75,1.45)(1.4,1.2)}
\rput(-.6,0.5){\makebox(0,0)[cb]{$Z_1$}}
\rput(0.6,0.1){\makebox(0,0)[lb]{$Z_2$}}
\rput(-.2,-1){\makebox(0,0)[cb]{$Z_3$}}
\rput(1.5,1.4){\makebox(0,0)[lc]{$R_1$}}
\rput(0.1,-1.8){\makebox(0,0)[ct]{$R_2$}}
\rput(-2,0.7){\makebox(0,0)[cb]{$R_3$}}
\end{pspicture}
}
\caption{\small Paths for arcs from the dual triangulation.}
\label{fi:ll}
\end{figure}
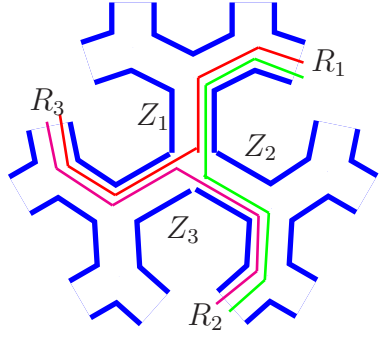

The {\bf proof} is the direct calculation. We begin with the pattern corresponding to a three-valent vertex in Fig.~\ref{fi:ll}. $Z_i$ are variables of three edges incident to the vertex and $R_i$ are the rest of variables along the corresponding line (note that all of variables in each of $R_i$ are incident to one and the same boundary component). Expression in (\ref{WP-SS}) then results in
$$
\sum_{i=1,2,3\mod 3} d(R_i+Z_i)\wedge d(R_{i+1}+Z_{i+1}).
$$
Consider now all expressions containing, say, $R_1$ and $R_2$. We have exactly two: one is presented above, another comes from the second vertex (immediately to the bottom-right from the first vertex in the figure; in this second vertex the relevant term will be $dR_2\wedge (dR_1+dZ_1+dZ_2)$ and the sum of these two terms just contributes $d(R_1+Z_1)\wedge dZ_2+dR_2\wedge dZ_2$, i.e., exactly the part of expression for $\Omega_1$ and $\Omega_2$ in Theorem~\ref{th:Omega}. Taking the sum over all edges (including pending edges but excluding loops), we obtain the theorem statement. 

\begin{remark}
Very recently, Bertola and Korotkin obtained an analogous formula (formula (6.5) in \cite{BK}) from Alekseev and Malkin \cite{AM} symplectic form. There symplectic form $\Omega_j$ (\ref{star}) was associated with $j$th cusp (hole of zero perimeter) restricted to a submanifold $\sum_{\beta\in I_j}dY_\beta=0$; this restriction allows choosing a ciliation (determining a linear, not cyclic, ordering) arbitrarily, but it also makes $dY_\beta$ dependent.
\end{remark}

\begin{figure}[tb]
%\hspace*{2cm}
%\epsfysize=6cm
%\vskip .2in
{\psset{unit=1}
\begin{pspicture}(-2,-1)(3,2)
\newcommand{\TADPOLE}{%
{\psset{unit=1}
\psline[linewidth=18pt,linecolor=blue](0,0)(0,1)
\pscircle[linewidth=2pt,linecolor=blue,fillstyle=solid,fillcolor=white](0,1.4){0.5}
\pscircle[linewidth=2pt,linecolor=blue,fillstyle=solid,fillcolor=white](0,1.4){0.2}
\psline[linewidth=14pt,linecolor=white](0,-0.2)(0,1.2)
}
}
\newcommand{\TADPOLELINE}{%
{\psset{unit=1}
\psarc[linecolor=red](-0.35,0.1){0.2}{270}{360}
\psline[linewidth=1pt,linecolor=red]{->}(-0.15,0.1)(-0.15,1.1)
\psline[linewidth=1pt,linecolor=red]{->}(0.15,1.1)(0.15,0.1)
\psarc[linecolor=red](0,1.4){0.35}{-63}{243}
}
}
\psline[linewidth=18pt,linecolor=blue](-2,0)(2.25,0)
\psline[linewidth=14pt,linecolor=white](-2.1,0)(2.35,0)
\rput(-1,0.25){\TADPOLE}
\rput(-1.07,0.25){\TADPOLELINE
\psline[linewidth=1pt,linecolor=red]{->}(-0.85,-0.1)(-0.35,-0.1)
\psarc[linecolor=red](0.35,0.1){0.2}{180}{270}
\psline[linewidth=1pt,linecolor=red]{->}(0.35,-0.1)(1.25,-0.1)
}
\rput(0.5,0.25){\TADPOLE}
\rput(0.43,0.25){\TADPOLELINE
\psarc[linecolor=red](0.35,0.1){0.2}{180}{270}
\psline[linewidth=1pt,linecolor=red]{->}(0.35,-0.1)(1.25,-0.1)
}
\rput(2,0.25){\TADPOLE}
\rput(2,0.25){\TADPOLELINE}
\rput{270}(2.25,-0.06){\TADPOLE}
\rput{270}(2.2,-0.06){\TADPOLELINE}
\psline[linewidth=1pt,linecolor=red]{->}(2.3,-0.15)(0.7,-0.15)
\psline[linewidth=1pt,linecolor=red]{->}(0.7,-0.15)(-0.8,-0.15)
\psline[linewidth=1pt,linecolor=red]{->}(-0.8,-0.15)(-2,-0.15)
\rput(-1.5,-0.5){\makebox(0,0)[ct]{$\pi$}}
\rput(-.25,-0.4){\makebox(0,0)[ct]{$b_1$}}
\rput(1.25,-0.4){\makebox(0,0)[ct]{$b_2$}}
\rput(2.75,-0.4){\makebox(0,0)[ct]{$b_3$}}
\rput(-1.45,0.8){\makebox(0,0)[rc]{$a_1$}}
\rput(0.05,0.8){\makebox(0,0)[rc]{$a_2$}}
\rput(1.55,0.8){\makebox(0,0)[rc]{$a_3$}}
\rput(-1.6,1.65){\makebox(0,0)[rc]{$w_1$}}
\rput(-0.1,1.65){\makebox(0,0)[rc]{$w_2$}}
\rput(1.4,1.65){\makebox(0,0)[rc]{$w_3$}}
\rput(3.6,.65){\makebox(0,0)[cb]{$w_4$}}
\end{pspicture}
}
\caption{\small A fat graph that is a spine of $\Sigma_{0,5,1}$.}
\label{fi:051}
\end{figure}
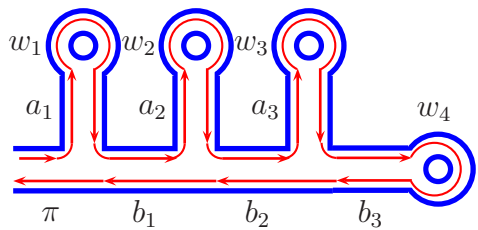

We conclude by the statement that the symplectic form in Theorem \ref{th:Omega} is inverse to the standard Poisson bracket (\ref{WP-PB}) on the set of extended shear coordinates. The proof is based on the mapping class group invariance of both $w_{\text{WP}}$ and $\Omega_{\text{WP}}$, which enables performing calculations for a graph most suitable for a given problem thus reducing solution to probing a finite number of cases. We treat in details only the case of $\Sigma_{0,s+1,1}$, i.e., the disc with $s$ holes without boundary cusps in the interior and with the single hole with the bordered cusp in the exterior (see the example of the corresponding fat graph for $s=4$ in Fig.~\ref{fi:051}). It has just one window and the order of edges (with a marking indicated in the figure) is $\pi\to a_1\to a_1\to b_1\to a_2\to a_2\to b_2\to a_3\to a_3\to b_3\to b_3\to b_2\to b_1\to \pi$ (recall that coefficients $w_i$ do not contribute to the symplectic form). We see that $d\pi$ drops out of the symplectic form $\Omega_{\text{WP}}$ whereas the six remaining variables collected into the vector $\mathbf v$ produce the form
$$
\Omega_{\text{WP}}=4\mathbf v\wedge \Omega \mathbf v^{\text{T}},\hbox{ where } \Omega=
\left[
\begin{array}{rrrrrr}
0  & 1  & 1 & 1 & 1 & 1  \\
-1  & 0  & 0 & 0 & 0 & 0  \\
-1  & 0  & 0 & 1 & 1 & 1  \\
-1  & 0  & -1 & 0 & 0 & 0  \\
-1  & 0  & -1 & 0 & 0 & 1  \\
-1  & 0  & -1 & 0 & -1 & 0  
\end{array}
\right]\hbox{ and }\mathbf v^{\text{T}}=\left[
\begin{array}{r}
da_1   \\
db_1  \\
da_2   \\
db_2   \\
da_3   \\
db_3  
\end{array}
\right].
$$
The corresponding Poisson bi-vector $\omega_{\text{WP}}$ restricted to the vector fields $\partial_{a_i}$, $\partial_{b_i}$ reads
$$
\omega_{\text{WP}}=\mathbf u\wedge \omega \mathbf u^{\text{T}},\hbox{ where } \omega=
\left[
\begin{array}{rrrrrr}
0  & 1  & 0 & 0 & 0 & 0  \\
-1  & 0  & 1 & -1 & 0 & 0  \\
0  & -1  & 0 & 1 & 0 & 0  \\
0  & 1 & -1 & 0 & 1 & -1  \\
0  & 0  & 0 & -1 & 0 & 1  \\
0  & 0  & 0 & 1 & -1 & 0  
\end{array}
\right]\hbox{ and }\mathbf u^{\text{T}}=\left[
\begin{array}{r}
\partial_{a_1}   \\
\partial_{b_1}  \\
\partial_{a_2}   \\
\partial_{b_2}   \\
\partial_{a_3}   \\
\partial_{b_3}  
\end{array}
\right],
$$
and it is easy to find that $\Omega\omega=-4\mathbb E$. 

\end{document}